\documentclass[twocolumn,superscriptaddress,nofootinbib,floatfix,aps,prc,10pt]{revtex4-2}
\usepackage{graphicx} 

\usepackage{amsmath} 
\usepackage{tabularx}
\usepackage{color} 
\usepackage{array} 
\usepackage{bm}
\usepackage{physics}

\usepackage{xspace}

\newcommand\lcmin{LC $\chi^2$-minimization method\xspace}

\begin{document}
\title{Towards shell model interactions with credible uncertainties}
\author{Oliver C. Gorton}
\email{gorton3@llnl.gov}
\affiliation{Lawrence Livermore National Laboratory, P.O. Box 808, L-414,
Livermore, California 94551, USA}
\author{Konstantinos Kravvaris}
\email{kravvaris1@llnl.gov}
\affiliation{Lawrence Livermore National Laboratory, P.O. Box 808, L-414,
Livermore, California 94551, USA}

\begin{abstract}
\edef\oldrightskip{\the\rightskip}
\begin{description}
\rightskip\oldrightskip\relax

\item[Background] The nuclear shell model is a powerful framework for predicting
nuclear structure observables, but relies on interaction matrix elements fit to
experimental data as its inputs. Extending the shell model’s applicability,
particularly toward dripline nuclei, requires efficient fitting methods and
credible uncertainty quantification. Traditional approaches face computational
challenges and may underestimate uncertainties.

\item[Purpose] We develop and test a framework combining eigenvector
continuation and Markov Chain Monte Carlo to efficiently fit shell model
interaction matrix elements and quantify their uncertainties.

\item[Methods] Eigenvector continuation is used to emulate shell model
calculations, reducing computational costs. The emulator enables Markov chain
Monte Carlo sampling to optimize interaction matrix elements and rigorously
assess parametric uncertainties. The framework is benchmarked using the USDB
interaction in the \textit{sd}-shell.

\item[Results] The emulator reproduces the USDB interaction with negligible
error, validating its use in shell model fitting applications. However, we find
that to obtain credible predictive intervals, the model defect of the shell
model itself, rather than experimental or emulator error, must be taken into
account in order to obtain credible uncertainties.

\item[Conclusions] The proposed framework provides an efficient and rigorous
approach for fitting shell model interactions and quantifying uncertainties.
Further, the normality assumption used in the past appears sufficient to
describe the distribution of interaction matrix elements. However, it is crucial
to account for model correlations to avoid underestimating uncertainties.

\end{description}
\end{abstract}

\maketitle

\section{Introduction}
\label{sec:intro}

For the past 75 years, the shell model (SM)~\cite{mayer1949on, haxel1949on,
caurier2005shell, nowacki2021neutron} has served as a foundational tool for
accurately predicting nuclear structure observables. These include binding
energies, spectra, electromagnetic and weak decays, as well as level density
information, which can all be used to inform applications in nuclear
astrophysics and nuclear technologies. The SM's success has led to the
development of various approaches aimed at extending its applicability, not only
to unexplored regions of the nuclear chart but also to the study of collective
phenomena~\cite{otsuka2001monte, dao2022nuclear}, a prospect that would naively
go against the single-particle understanding that the SM offers.

At its core, the SM describes the nucleus using a handful of degrees of freedom,
namely valence protons and neutrons in a mean field, interacting with
two-particle forces. How exactly the many-body problem that includes
nucleon-nucleon and many-nucleon interactions can be reduced to such a
description is a matter of active research~\cite{hjorth-jensen1995realistic,
stroberg2019nonempirical, miyagi2020abinitio, coraggio2020perturbative,
dikmen2015abinitio, sun2018shellmodel}. Despite these first-principles efforts,
the most accurate SM predictions typically rely on inputs derived from direct
fits to experimental data~\cite{brown1988status, honma2002effective,
brown2006new, utsuno2012shell, caurier2014merging, lubna2020evolution,
magilligan2020new}. 

The computational cost of these fits has inhibited rigorous error analysis.
Recent progress has enabled rudimentary uncertainty quantification (UQ)
approaches~\cite{fox2020uncertainty, fox2023uncertainty,
yoshida2018uncertainty}, though these are currently restricted to relatively
small-scale calculations. Consequently, there is a pressing need to develop a
framework that can efficiently fit phenomenological SM interactions to
experimental data while simultaneously providing credible uncertainties.

Fitting shell model interactions directly translates to fitting the mean-field
energies and residual two-body matrix elements. One first defines the valence
(or active) space of single particle levels that the valence nucleons can
occupy. The effective coupling constants of the interaction, collectively
labeled by $\bm x$, include the single-particle energies (SPEs) and two-body
matrix elements (TBMEs) of and between the SM orbitals. The total interaction
with one- and two-body terms can be written
as~\cite{brown2006new}:
\begin{equation}\label{eq:int}
    \hat H(\bm x) = \sum_i \epsilon_i \hat n_i
    + \sum_{i\le j, k\le l; JT} V_{ijkl; JT} 
    \hat T_{ijkl; JT}
\end{equation}
where the indices $i, j, k, l$ label the single-particle orbits (typically
harmonic oscillator states); the collective index $i$ is short for all quantum
numbers defining an orbit $(n_i, l_i, j_i)$, with principle quantum numbers
$n_i$, orbital angular momentum $l_i$, and total angular momentum $j_i$. The
number operator for a given shell $i$ is represented by $\hat{n}_i$ and
$\hat{T}$ is the scalar two-body density operator~\cite{brown2006new}. 

The interaction matrix elements $\bm x$ are shorthand for the full set of SPEs
and TBMEs: $x: (\epsilon_i, V_{ijkl;JT})$, which must be obtained from either
theoretical or experimental constrains. For a given parametrization of the
interaction, $\bm x$, the eigenfunctions $\ket{\Psi(\bm x)}$ and eigenenergies
$E(\bm x)$ are found by solving the time-independent Schrodinger equation:
\begin{equation}\label{eq:se}
    \hat{H}(\bm x) \ket{\Psi(\bm x)} = E(\bm x) \ket{\Psi(\bm x)}.
\end{equation}
In the full configuration interaction (FCI) framework with a frozen core,
equation~\eqref{eq:int} is typically restricted to a finite basis of Slater
determinants. This approach represents the Hamiltonian as a matrix, whose
eigenpairs correspond to the nuclear wave functions and energies. 

SM interactions, such as the USD-family (USD~\cite{wildenthal1984empirical,
brown1988status}, USDA/B~\cite{brown2006new}, USDC~\cite{magilligan2020new})
have been highly successful in reproducing experimental binding energies,
low-lying excitation energies, and transition probabilities. For
\textit{sd}-shell nuclei, the most common variants, USDA and USDB, consist of
three single-particle energies and 63 two-body matrix elements. These matrix
elements were empirically fit with specific linear combinations of the 66
parameters iteratively updated to minimize a $\chi^2$ fit to 608 energy levels
across 77 nuclei~\cite{brown2006new}. This procedure is remarkably efficient,
requiring around 30 iterations, with each iteration taking 12 hours. It achieved
a local optimum while keeping computational demands manageable, but provided
limited insight into the robustness of the predictions.

The goals of this paper are: 1) to benchmark a new, computationally efficient
methodology for fitting SM interactions, which is critical for generating
uncertainty-quantified interactions beyond the \textit{sd}-shell--particularly
in neutron-rich model spaces that are essential for studying r-process
nucleosynthesis; and 2) to rigorously assess the parametric uncertainty and
covariances of the phenomenological SM. While previous studies have investigated
the uncertainty of the USDB interaction~\cite{fox2020uncertainty,
fox2023uncertainty} and the SM more broadly~\cite{yoshida2018uncertainty}, these
efforts have been limited to approximate statistical methods due to
computational costs. 

Recent developments in computational methods offer promising solutions to these
goals. Eigenvector continuation (EC) has emerged as a powerful tool for
emulating solutions to parametric Hamiltonians. By constructing a
reduced basis from known eigenstates, EC circumvents the need for repeated
diagonalization of large Hamiltonian matrices, dramatically reducing
computational costs. This acceleration makes advanced fitting algorithms, such
as Markov Chain Monte Carlo (MCMC), feasible for fitting SM interactions. MCMC
also provides a rigorous framework for exploring parameter uncertainties in
nonlinear models, overcoming the limitations of traditional $\chi^2$
minimization techniques.

In this paper, we develop and test a framework combining EC and MCMC to
efficiently fit SM interactions and obtain credible uncertainties. Using the
$sd$-shell as a benchmark, we demonstrate that EC can reproduce the well-known
USDB interaction with negligible error, validating its use as an emulator for SM
calculations. (See Section~\ref{sec:methods} and Section~\ref{sec:LC}.) We then
employ MCMC to generate uncertainty-quantified interactions, investigating
concerns of nonlinearities and correlations in the parameter space that were
previously neglected. (See Section~\ref{sec:mcmc}.) Finally, we propose a
practical approach to correct these underestimations, ensuring that predictive
intervals better reflect the true uncertainty of SM calculations.

\section{Methods}
\label{sec:methods}

In this section, we review some aspects of the FCI SM and describe the
algorithms employed to fit the interaction matrix elements. Next, we briefly
introduce the specific eigenvector continuation (EC) approach used to emulate
the FCI SM, including details for generating the reduced basis. Finally, we
discuss how we validated the accuracy of the EC model as a practical emulator
for the FCI SM, with a preview of how well the EC emulator can be used to fit SM
interactions.

The form of our parametric Hamiltonian $H(\bm x)$ is that of the interacting SM
given in equation~\eqref{eq:int}. The interaction matrix elements
$\bm x$ are shorthand for the full set of SPEs and TBMEs: $x: (\epsilon_i,
V_{ijkl;JT})$, which are to be constrained by experimental measurements of
spectroscopic observables. As a further shorthand, we can consider the matrix
$H(\bm x)$ as a sum of matrices ${O}_n$ with coefficients $x_n$ from each term
in~\eqref{eq:int}:
\begin{equation}\label{eq:hsum}
    H(\bm x) = \sum_n x_n O_n,
\end{equation}
where each $O_n$ is one of the $\hat n_i$ or $\hat T_{ijkl; JT}$ operators cast
as a matrix: $O_{n,ij} = \langle \phi_i | O_n | \phi_j \rangle$, given some
complete set of basis states $\ket{\phi_i}$. We take the convention that the
coefficients $x_n$ carry the units of MeV and the operator matrix elements are
dimensionless. With mass-dependent two-body matrix elements (TBMEs), we redefine
the coefficients:
\begin{equation}
    H(\bm x) = \sum_{i\in \mathrm{SPE}} x_i O_i 
    + \sum_{i\in \mathrm{TBME}} x_i \left( \frac{A_0}{A}\right)^p O_i,
\end{equation}
where $A$ is the mass of the nucleus, $p = 0.3$, and $A_0=18$ for the
$sd$-shell~\cite{brown2006new}. With this factorization, eigenvalues of the
Hamiltonian ($\bm \lambda)$ for the $k$th energy level ($\lambda_k$) can be
written as:
\begin{equation}\label{eq:lambda}
    \lambda_k = \sum_i x_i \tilde{\beta}_i^k,
\end{equation}
where,
\begin{equation}
    \tilde{\beta}_i^k(A) = 
    \begin{cases}
        \langle \Psi_k(\bm x) | O_i | \Psi_k(\bm x) \rangle, i \in \text{SPE} \\
        \left( \frac{A_0}{A}\right)^p 
        \langle \Psi_k(\bm x) | O_i | \Psi_k(\bm x) \rangle, i \in \text{TBME}
    \end{cases},
\end{equation}
computed using the $k$-th eigenvectors $\ket{\Psi_k}$ of the
Hamiltonian.

As in Ref.~\cite{brown2006new}, the eigenvalues of the Hamiltonian in
equation~\eqref{eq:lambda} can be related to the experimental nuclear binding
energies by:
\begin{equation}\label{eq:BE}
    BE = \lambda_k + BE(^{16}\text{O}) + E_C(Z),
\end{equation}
where $BE(^{16}\text{O})$ is the binding energy of the $^{16}\text{O}$ nucleus,
and $E_C(Z)$ is a correction for the Coulomb interaction of the valence protons.
We take the $E_C$ values listed in Ref.~\cite{brown2006new} (originally from
Ref.~\cite{chung1976empirical}). Following the decision of Brown and Richter, we
fix the ground state energies to be given by the eigenvalues $\lambda_k$, and
compare to the experimental ground state binding energies after subtracting the
core $(BE(^{16}\text{O}))$ and Coulomb correction $(E_C(Z))$ energies.
Meanwhile, the excited states are fitted directly as excitation energies
$\lambda_k - \lambda_0$, where $\lambda_0$ is the ground state energy of a given
nucleus.

\subsection{Fitting procedures}\label{sec:fitting}

In order to constrain the interaction matrix elements $\bm x$ using experimental
data, we must define a fitting procedure. We adopt two approaches.

The first approach employs Monte Carlo techniques to estimate the probability
distribution of the SM interaction parameters, incorporating statistical
constraints from experimental data. The second approach replicates earlier
methods based on $\chi^2$ minimization, which serves as a simplified and
approximate alternative to the Monte Carlo method. 

A key advantage of the Monte Carlo approach is its ability to sample arbitrarily
complex probability distributions arising from the mapping between the model
observables, $\bm{\lambda}$, and the model parameters, $\bm{x}$. Given the
non-invertible nature of the model and the large number of parameters involved,
we use Markov Chain Monte Carlo (MCMC) to solve this inverse problem. An
additional advantage of the MCMC approach is that it avoids the linear
approximation~\cite{fox2020uncertainty} typically used to compute derivatives of
the $\chi^2$ cost function. This linear approximation neglects the inherently
nonlinear relationship between the nuclear wavefunctions and the interaction
parameters, which could lead to inaccuracies in parameter estimation. By
contrast, MCMC fully accounts for these nonlinearities, providing a more
rigorous and flexible framework for constraining the interaction parameters.

To use Markov Chain Monte Carlo (MCMC) to fit SM interactions, we sample from a
probability distribution relating SM predictions to experimental data. For model
parameters $\bm x$, and given the experimental energies $\bm E$ and their
covariances $\Sigma$, we define a posterior probability distribution as:
\begin{equation}\label{eq:bayes}
    P(\bm x | \bm E, \Sigma) 
    = \frac{L(\bm E, \Sigma | \bm x) p(\bm x)}
    {\int L(\bm E, \Sigma | \bm x) p(\bm x) d^n \bm x},
\end{equation}
which is just Bayes' theorem~\cite{gelman1995bayesian}. The prior probability
density function for the interaction matrix elements $(p(\bm x))$ can be
obtained from either an effective interaction derived from a more fundamental
nuclear force, or from a prior fit. MCMC allows us to directly sample from
$P(\bm x | \bm E, \Sigma)$ without computing the denominator of the right hand
side of equation~\eqref{eq:bayes}, which would in turn require a rather
impractical sampling over \textit{all} possible values of $\bm x$.

The likelihood function $L(\bm E, \Sigma | \bm x)$ should describe the
probability of observing the set of energies with mean $\bm  E$ and covariance
$\Sigma$, given a set of parameters $\bm x$. Designing a likelihood function
requires some knowledge of the measurements of $(\bm  E, \Sigma)$, and the
expected source of differences between the model and the measurements. We used a
standard log-likelihood function~\cite{gelman1995bayesian}, given by the natural
logarithm of a multivariate normal likelihood function:
\begin{equation}\label{eq:logp}
\ln L(\bm E, \Sigma | \bm x) =
    - \frac{1}{2}(\bm{r}^T \Sigma^{-1} \bm{r} + \ln |\Sigma| + k\ln 2\pi),
\end{equation}
where $\bm r = \bm \lambda(\bm x) - \bm E$ is the residual between the
experimental energies $\bm E$ and the model prediction $\bm \lambda(\bm x)$. $k$
is the number of matrix elements in $\bm x$, i.e. the number of free parameters.
The data, given by $\bm E$ with the associated covariance matrix
$\Sigma_\text{exp.}$, includes all energy levels and binding energies across the
nuclei considered. 

The covariance matrix models the expected residual given what we know about the
source of the errors. Here, the covariance contains contributions from both the
reported experimental covariance and the theoretical covariance
($\Sigma_\text{th.}$):
\begin{equation}\label{eq:cov}
    \Sigma = \Sigma_\text{exp.} + \Sigma_\text{th.},
\end{equation} 
both of which are assumed to be diagonal (uncorrelated). Note that in the case
of fitting energy levels with the SM, the experimental uncertainty is
vanishingly small (at the level of a single~keV), especially when compared with
the expected theoretical error ($>100$~keV), so
$\Sigma\approx\Sigma_\mathrm{th.}$

This assumption of uncorrelated energy measurements should be valid for
unrelated experimental measurements. (And, unfortunately, even nominally
correlated measurements are seldom reported with a correlation analysis.) On the
other hand, we know that the assumption of a diagonal theoretical covariance
matrix is flawed: the model outputs $\bm \lambda(\bm x)$ are not random and
depend on parameters which are fewer in number than the model outputs.
Furthermore, depending on the nature of the model defect (the physics missing
from the model), the assumption of a multivariate normal probability for the
residuals also comes into question. Our choice for $\Sigma_\text{th.}$ to
compensate for these unaccounted for correlations will be discussed in detail in
Section~\ref{sec:underest}.

\subsubsection{Chi-squared minimization}\label{sec:lcmethod}

The $\chi^2$ minimization algorithm used in~\cite{brown2006new} can be
interpreted as an approximate method to obtain the maximum likelihood estimate
of the log-likelihood function, Eq.~\eqref{eq:logp}. The maximum of the
log-likelihood coincides with the minimum of the $\chi^2$ cost function:
$\chi^2 = \bm r^T \Sigma \bm r$. By assuming the covariance matrix $\Sigma$ is
diagonal ($\Sigma_{ij}=\sigma^2_i\delta_{ij}$), we obtain the optimality
condition:
\begin{align}\label{eq:min1}
    0 = \sum_{n=1}^{N_\text{data}} \frac{\lambda_n(\bm x) - E_n}{\sigma_n^2} 
    \frac{\partial \lambda_n(\bm x)}{\partial x_j},
\end{align}
which is solved by some optimal set of matrix elements $\bm x^*$, to be
determined. To evaluate the derivatives of the energies, one uses the
Feymann-Helmann (FH) theorem, which applies whenever the wavefunction
$|\Psi_n(\bm x)\rangle$ is an eigenstate of the Hamiltonian $H(\bm x)$ with
matching parameters $\bm x$:
\begin{equation}\label{eq:FH}
    \frac{\partial \lambda_n}{\partial x_j} 
    =
    \left \langle
    \Psi_n(\bm x) \left |
    \frac{\partial H(\bm x)}{\partial x_j}
    \right |\Psi_n (\bm x)
    \right \rangle
    = \tilde{\beta}_j^n.
\end{equation}
Because we begin with an ansatz for the parameters $\bm x^{(0)}$ and
wavefunctions $|\Psi_i(\bm x^{(0)})\rangle$, there is a mismatch between the
parameters of the wavefunction and those of the derivative, and the FH theorem
is violated. We therefore must either assume that the wavefunctions depend only
weakly on the parameters $\bm x$ (the \textit{linear approximation} described
in~\cite{fox2020uncertainty}), or we must iterate until convergence is reached.
Substituting Eq.~\eqref{eq:lambda} for $\lambda_k$ and using Eq.~\eqref{eq:FH}
in Eq.~\eqref{eq:min1}, one obtains a system of linear equations: 
\begin{equation}\label{eq:lineqn}
    G \bm x = \bm e,
\end{equation}
where the ``error matrix'' $G$ is:
\begin{equation}
    G_{ij} \equiv \sum_n \frac{\tilde{\beta}_i^n \tilde{\beta}_j^n}{\sigma_n^2},
\end{equation}
having units of 1/MeV$^2$, and 
\begin{equation}
    e_i \equiv \sum_n \frac{E_n \tilde{\beta}_i^n}{\sigma_n^2},
\end{equation}
having units of 1/MeV. By solving Eq.~\eqref{eq:lineqn}, we obtain an
approximate solution to the optimal set of matrix elements: $\bm x^{(i+1)} =
G^{-1}(\bm x^{(i)}) \bm e(\bm x^{(i)})$. We iterate several times, updating the
wavefunctions $|\Psi_n(\bm x^{(i)}) \rangle$ used to compute the
$\tilde{\beta}$s in $G$ and $\bm e$, until $\bm x^{(i+1)} = \bm x^{(i)}$.

\subsubsection{Linear combinations method}\label{sec:lccmethod}

We also make use of the linear combinations method, which is also called the
singular value decomposition (SVD) or principal component analysis (PCA) method.
The inverse of the error matrix $G$ approximates the covariance of the optimal
parameters $\bm x^*$, $G^{-1}\approx \Sigma_x$~\cite{brown2006new,
fox2020uncertainty}. To identify which linear combinations of matrix elements
are most constrained by $\chi^2$ minimization, we can diagonalize the error
matrix $G$, $D = A G A^T$, using a linear transformation matrix $A$. We can then
define linear combinations (LCs) $\bm y = A \bm x$ and $\bm c = A \bm e$ to
obtain a minimization equation in the space of $A$: $D \bm y = \bm c$. The
inverse of the diagonal error matrix $D$ then approximates the covariance of the
transformed parameters $\bm y$: $D^{-1}\approx \Sigma_y$. 

The LCs with the smallest(largest) $\Sigma_y$ are the most(least) constrained.
Following Ref.~\cite{brown2006new}, we update only the first $N_d$ most
constrained linear combinations during each iteration of the $\chi^2$
minimization algorithm, and the remaining linear combinations are set to some
preferred set of matrix elements. In the case of Ref.~\cite{brown2006new}, $\bm
x_0$ are the RGSD matrix elements of Ref.~\cite{hjorth-jensen1995realistic}. In
the language of Eq.~\eqref{eq:bayes}, the linear combinations method amounts to
using a completely uninformative prior for the first $N_d$ LCs and using an
infinitely narrow prior for the remaining least constrained LCs (although, the
definition of the least constrained LCs depends on the choice of ``prior'').

\subsection{Eigenvector continuation}\label{sec:ec}

Eigenvector continuation (EC) is a computational method that approximates
solutions to the Schrödinger equation for parametric Hamiltonians $H(\bm x)$ by
leveraging a reduced basis constructed from known eigenstates. This approach
enables rapid emulation of the full SM without repeated diagonalization of large
Hamiltonian matrices. The original application of EC in nuclear physics was to
compute $H(\bm x)$ for a physical value of $\bm x$ which is otherwise
computationally prohibitive~\cite{frame2018eigenvector}. It was quickly realized
that EC could be used as a tool for parameter calibration and uncertainty
quantification~\cite{konig2020eigenvector, yoshida2022constructing} by
approximating full re-diagonalization after each small perturbation of the
interaction matrix elements. In this work, we apply EC to emulate the SM and
accelerate a global fitting procedure of the interaction matrix elements.

The basic equations for EC, given a parametric Hamiltonian of the
form~\eqref{eq:hsum} with an M-scheme dimension $d$, are as follows. 
We construct the columns of a
projection matrix $V_{d\times q}$ using $q$ vectors of dimension $d$ (with $q\ll
d$). We denote this set of vectors $\{\bm v_i\}_{i=1,...,q}$, which form the EC
basis (to be defined). We use $V_{d\times q}$ to project each operator
in~\eqref{eq:hsum} into the EC basis: $O_{n; q\times q} = V^T_{q\times d} O_{n;
d\times d} V_{d\times q}$. Then, the EC Hamiltonian matrix for any new
interaction $\bm x^\text{new}$ is simply $T_{q\times q}(\bm x^\text{new}) =
\sum_{n=1}^k x^\text{new}_n O_{n; q\times q}$, which has approximately the same
eigenvalues as $H_{d\times d}(\bm x^\text{new})$. In subsection~\ref{sec:basis}
we detail how the EC basis is constructed, and in~\ref{sec:validation} we show
the quality of the approximation.

\subsubsection{EC basis construction}\label{sec:basis}

The first step in EC is to construct a reduced basis $\{\bm
v_i\}_{i=1,...,q}$ that spans the space where new solutions will be sampled
from. This basis is generated by solving the Schrödinger equation
for a set of $N$ random trial interactions $\bm x$, chosen near the target
region. For each trial interaction, we compute the low-lying eigenstates of the
Hamiltonian $H(\bm x)$, which are the vectors $\bm v_i$ used to
form the EC basis. The basis vectors are orthonormalized to simplify subsequent
diagonalization steps and ensure numerical
stability~\cite{sarkar2022eigenvector}.

In this study, the trial interactions $\bm x$ were sampled from a uniform
distribution centered on the USD interaction matrix elements
$\bm{x}_\text{USD}$~\cite{wildenthal1984empirical}, with a relative uncertainty
of 40\%. This choice corresponds to a root-mean-squared (RMS)
deviation of approximately 450~keV between sampled interactions and USD,
comparable to the differences among the RGSD, USD, USDA, and USDB
interactions~\cite{brown2006new}. To clarify further comparisons, we will define
the ``interaction root-mean-squared deviation'' (IRMS) as the RMS deviation
between two sets of interaction matrix elements:
\begin{equation}
    \text{IRMS}(\bm x, \bm x') = \sqrt{
        \frac{1}{k} \left( \sum_{i=1}^{k} (x_i - x_i')^2 \right)
    },
\end{equation}
where $k$ is the sum of the number of single-particle energies and TBMEs, and
$x_i$ and $x_i'$ are interaction matrix elements being compared. For the
\textit{sd}-shell, $k=66$.

If the new minimum obtained during the fit were significantly far from the
training region, it may have be necessary to update the emulator by computing
new full samples near the new minimum. We did not find resampling necessary
here.

For each trial interaction, we generated eigenstates for all nuclei, total
angular momentum $J$, and isospin $T$ combinations in the training dataset. We
created and partially diagonalized all Hamiltonians required to obtain $m$
(lowest-lying) states for each combination. In the present application, the
dataset includes all experimental levels to which we are fitting from
Ref.~\cite{brown2006new}. The corresponding file is supplied the Supplemental 
Material~\cite{supp}.

To maximize computational efficiency, any new basis vector with a vector overlap
exceeding 0.99 with existing basis vectors is rejected. This ensures the basis
remains compact while retaining sufficient accuracy. Basis
construction halts if the number of sampled eigenstates $q=N\times m$ reaches
the full model space dimension $d$, where $m$ is the number of low-lying
eigenstates per nucleus per $J, T$ combination.

For this study, we found that a basis size of $N=40$ trial interactions, each
contributing $m=10$ eigenstates, offered sufficient accuracy while maintaining
modest computational requirements. This configuration required approximately
5~seconds of runtime (serial) to evaluate all 608 energy levels across 77
nuclei, with a memory footprint of 15~GB. Table~\ref{tab:models} summarizes the
performance of EC models with varying basis sizes.

\subsubsection{Emulator validation}\label{sec:validation}

\begin{figure}[ht]
    \centering
    \includegraphics[width=\linewidth]{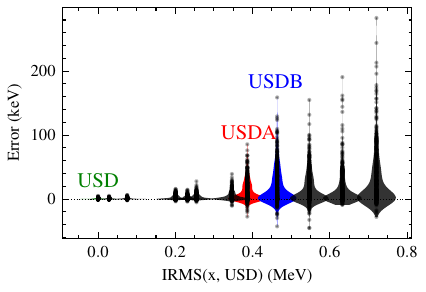}
    \caption{Data root-mean-squared error (DRMS) of the EC emulator compared to
    the exact SM values for the 77 binding energies and 531 excitation energies,
    and as a function of the IRMS of the test interaction compared to USD. (For
    comparison, the average IRMS of the training-point interactions is
    0.45~MeV.) For each of the 12 test interactions, we plot a
    point for the error of each of the 608 data points, and on top of this we
    draw a violin plot where the width shows the density of the 608 points.}
    \label{fig:ecvalidation} 
\end{figure}

To validate the EC emulator as a practical substitute for the full SM, we tested
its accuracy against exact SM calculations for a set of randomly chosen
interactions. These ``test interactions'' were selected to span increasing IRMS
values relative to USD, allowing us to assess the emulator's performance as the
test interaction deviates from the training domain. Fig.~\ref{fig:ecvalidation}
illustrates the relationship between the IRMS of the test interaction and the
data root-mean-squared deviation (DRMS) of the emulator's predictions:
\begin{equation}
    \text{DRMS} = \sqrt{\frac{1}{N_\text{BE}+N_\text{Ex.}} 
    \sum_{n=1}^{N_\text{BE}+N_\text{Ex.}} (\lambda_n - E_n)^2},
\end{equation}
where $N_\text{BE}=77$ is the number of binding energy data points,
$N_\text{Ex.}=531$ is the number of excitation energy data points, $E_n$ is the
experimental value, and $\lambda_n$ is the emulator's prediction.

The EC emulator exhibited negligible error for test interactions with IRMS
values up to 250~keV relative to USD. For interactions closer to the fitting
domain (e.g., USDA and USDB), the DRMS was less than 30~keV, with more than half
of errors less than 6~keV, significantly smaller than the expected theoretical
uncertainty of the untruncated SM ($> 150$~keV). Table~\ref{tab:ecvalidation}
summarizes the emulator's performance across 12 validation tests, including
comparisons to USDA and USDB interactions. We also report typical memory and
runtime requirements of each EC model and compare to the full SM
code~\cite{CoSMo}.
\begin{table}[ht!]
    \centering
    \begin{tabular}{l c c c c c}
        Test & IRMS(USD) & DRMS & 10-th & 50-th & 90-th \\
        \hline
        \hline
        USD    &   0 & 0.8 & -0.006 & 0.1 & 1\\
        Rnd. 1 &  28 & 0.8 & -0.005 & 0.1 & 1\\
        Rnd. 2 &  75 & 1 & -0.007 & 0.1 & 2\\
        Rnd. 3 & 200 & 4 & -0.005 & 0.4 & 7\\
        Rnd. 4 & 230 & 3 & -0.007 & 0.4 & 5\\
        Rnd. 5 & 250 & 5 & -0.006 & 0.6 & 8\\
        Rnd. 6 & 350 & 7 & -0.005 & 0.9 & 13\\
        USDA   & 390 & 18 & -0.010 & 3.0  & 34\\
        USDB   & 460 & 32 & -0.010 & 5.4  & 56\\
        Rnd. 7 & 550 & 28 & -0.040 & 3.2  & 50\\
        Rnd. 8 & 630 & 28 & -0.008 & 1.7  & 43\\
        Rnd. 9 & 720 & 48 & -0.005 & 4.9  & 82 \\
        \hline
        \hline
    \end{tabular}
    \caption{Distributions of errors of the $N=40$, $m=40$ EC model for twelve
    test cases. We report the DRMS, and the 10-th, 50-th, and 90-th percentiles
    of error distributions shown in Fig.~\ref{fig:ecvalidation}. Units are in
    keV.}
    \label{tab:ecvalidation}
\end{table}

In general, the emulator's error is governed by two factors: 1) the relative
size of the EC basis compared to the full SM dimension; and 2) the distance of
the test interaction from the training domain. For the $N=40$, $m=10$ model, the
emulator error remains acceptably small ($<$10~keV), making it a reliable tool
for fitting new SM interactions.

We also characterized the impact of the EC model size on the LC $\chi^2$
minimization fitting procedure described in Section~\ref{sec:LC}, using the EC
model as a proxy for the FCI SM. For each EC model size, we computed the IRMS of
the fitted interaction compared to the USDB interaction. By this metric, the
$N=40$, $m=10$ model reaches an accuracy comparable to the exact SM ($\approx
12$~keV). Table~\ref{tab:models} shows the results for the other models tested
along with typical memory and runtime requirements of each EC model and compared
to the full SM code~\cite{CoSMo}.
\begin{table}[ht!]
    \centering
    \begin{tabular}{c c c c c}
        $N$ & $m$ & Memory (GB) & Runtime (s) & IRMS$(\bm x, \text{USDB})$ (keV) \\
        \hline
        \hline
        10  & 10 & 1.2 & 0.5 & 73\\
        20  & 10 & 4.1 & 1.5 & 39\\
        30  & 10 & 8.4 & 2.4 & 18\\
        40  & 10  & 15  & 5  & 12 \\
        \multicolumn{2}{c}{FCI} & - & $\approx$4000 & 12\\
        \hline
        \hline
    \end{tabular}
    \caption{Performance of EC models of different sizes including typical
    memory and total runtime requirements for EC models of different sizes. The
    runtime includes the time to evaluate all 608 energy levels of the 77
    $sd$-shell nuclei. The final column shows the IRMS relative to USDB of the
    \textit{sd}-shell interaction fitted using each model.}
    \label{tab:models}
\end{table}

In summary, the EC method provides an efficient and accurate means of emulating
the SM, significantly reducing computational demands while preserving predictive
power. In particular, we expect that within the domain of input interactions we
expect to encounter, the typical error introduced by the EC emulator on the
predicted energy levels is on the order of $< 10$~keV, which is an order of
magnitude smaller than the expected theoretical uncertainty of the SM. This
capability paves the way for uncertainty-quantified interaction fits in regions
of the nuclear chart where traditional SM calculations are computationally
prohibitive.

\section{Results}\label{sec:results}

The goal of any UQ approach is to reliably assess the expected error of model
predictions. In the following sections, we show that extending previous efforts
to quantify uncertainty in the phenomenological SM, particularly using the USDB
interaction, leads to a significant underestimation of uncertainty. We first
reproduce earlier methods based on least constrained minimization (\lcmin,
Section~\ref{sec:LC}), and then replace $\chi^2$ minimization with Markov Chain
Monte Carlo (MCMC, Section~\ref{sec:mcmc}). This extension explores a broader
parameter space and removes the assumption of multivariate normality for model
parameters. From this we find that earlier \lcmin methods are sufficient for
describing the distribution of matrix elements. On the other hand, empirical
tests show that both methods result in a systematic underestimation of
uncertainty by a factor of three. We propose correcting this underestimation by
increasing a key hyperparameter $\Sigma_\mathrm{th.}$, ensuring predictive
intervals better reflect the actual model error. To conclude the section, we
compare both energies and transition probabilities produced by the different
methods.

\subsection{LC $\chi^2$ minimization}\label{sec:LC}

We replicated the method of Brown and Richter~\cite{brown2006new} to reproduce
the USDB interaction, substituting the EC emulator for the full shell model
(SM). The EC emulator was configured with 40 samples and 10 levels per sample
(the $N=40$, $m=10$ model). This replication serves two purposes: first, to
validate the EC emulator's accuracy by reproducing earlier results, and second,
to recompute the most important linear combinations (LCs) of matrix elements,
enabling pre-optimization of the MCMC fitting procedure discussed in
Sec.~\ref{sec:mcmc}.

Following the conditions of Ref.~\cite{brown2006new}, we initialized the 66
matrix elements with USD values and used $(\Sigma_\text{th.}){ij} =
\sigma_\mathrm{th.}^2\delta_{ij}$, with $\sigma_\mathrm{th.}=100$~keV. For USDA
and USDB, we varied 30 or 56 LCs with the largest error-matrix eigenvalues,
respectively, while the remaining LCs were fixed to renormalized G-matrix
elements (RGSD) from Ref.~\cite{hjorth-jensen1995realistic}. The fit used the
same 77 ground-state energies and 531 excitation energies as
Ref.~\cite{brown2006new}, with ground-state energies corrected according to
Eq.~\eqref{eq:BE}, and excited states fitted as excitation energies relative to
the ground state.

Convergence of the model residuals and matrix elements was achieved within 30
iterations. For USDB, the DRMS value reached 133~keV, and the IRMS of the
LC-constrained matrix elements relative to USDB was 12~keV, consistent with
results using the full SM. The EC emulator introduced negligible errors, with a
DRMS discrepancy of only 3~keV compared to Ref.~\cite{brown2006new} and IRMS
differences on the order of tens of keV. These deviations are insignificant
compared to the typical expected error for $sd$-shell SM calculations (150~keV).
Numerical precision differences in diagonalization algorithms and matrix element
storage likely account for the small residuals relative to published USDA and
USDB interactions.

When we repeated the fit while varying 30 LCs ($N_d=30$), the converged DRMS
using the EC model was 175~keV, and the IRMS with respect to USDA was 10~keV. 
We also tested the sensitivity of the LC minimization to initial conditions by
varying the starting matrix elements across 22 replication studies. The matrix
elements and unitary transformation matrix $A$ consistently converged to the
same values within numerical precision, confirming the robustness of the method
to the initial matrix elments.

\begin{figure}[t!]
    \centering
    \includegraphics[width=\linewidth]{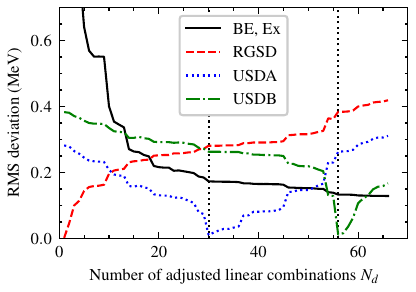}
    \caption{Different RMS metrics showing the behavior of the
    fitted interaction as we increase the number of adjusted linear
    combinations. We show IRMS of the fitted interaction relative to the RGSD,
    USDA, and USDB interactions, and the solid black line shows the DRMS of the
    predicted 77 binding energies and 531 excitation energies used in the fit.}
    \label{fig:rmsvsnd}
\end{figure}

We reproduced Figure 4 of Ref.~\cite{brown2006new} (here in
Fig.~\ref{fig:rmsvsnd}), which shows the DRMS and IRMS as functions of the
number of varied LCs ($N_d$). As expected, the IRMS relative to USDA and USDB
minimized at $N_d=30$ and $N_d=56$, respectively (see Fig.~\ref{fig:rmsvsnd}).
The high sensitivity of the IRMS is a direct result of the established
finding~\cite{brown2006new, fox2020uncertainty} that the constraints placed on
the LCs are concentrated to just a few leading terms (the terms best
constrained). However, we argue that rather than restricting the least
constrained parameters to some constant prior value (here, and in
Ref.~\cite{brown2006new}, the RGSD value), it is more appropriate to maintain
the value obtained from the fit and to report the estimated uncertainty.

\begin{figure}[ht!]
    \centering
    \includegraphics[width=\columnwidth]{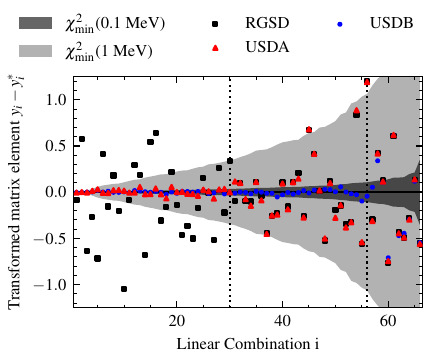}
    \caption{Transformed matrix elements $\bm y = A \bm x$ for $\bm x=$ RGSD,
    USDA, and USDB, relative to the optimal fit values $\bm y^*$ from the
    $N_d=66$ $\chi^2$-minimization. $A$ diagonalizes the error matrix of the
    fit. Verifies the relationship between the linear combinations USDA, USDB,
    and the RGSD matrix elements, and shows the relative uncertainty of the LCs
    as determined by the LC $\chi^2_\text{min}$ method using either
    $\sigma_{th}= 0.1$~MeV or $\sigma_{th}= 1.0$~MeV. The RGSD LCs are
    incompatible with \lcmin values given the uncertainties assuming
    $\sigma_{th}= 0.1$~MeV.} 
    \label{fig:ylc}
\end{figure}

In Fig.~\ref{fig:ylc}, we compare LC-transformed USDA and USDB matrix elements
to optimal values from our fit with $N_d=66$. We applied the linear
transformation $A$ (defining the optimal LCs for $N_d=66$) to transform the
RGSD, USDA, and USDB matrix elements, comparing each set to the optimal values
$y_i^*$. The plot highlights deviations from $y_i^*$, with the $1\sigma$
uncertainty region shaded based on $\sigma_i = \sqrt{1/D_{ii}}$, derived from
the diagonal error matrix $D$ (Sec.~\ref{sec:lccmethod}). The $\chi^2$
minimization used $\sigma_\mathrm{th.}=100$~keV, setting the absolute scale of
uncertainties, as experimental uncertainties are much smaller. Uncertainties
assuming $\sigma_\mathrm{th.}=1$~MeV are also shown.

Below $i=30$ for USDA and $i=56$ for USDB, the LCs of the published interactions
match our optimal fit values, while above these thresholds, they revert to RGSD
values, as defined. Most RGSD-transformed LCs deviate significantly from the
optimal values, exceeding 1$\sigma$ uncertainty bounds. To encompass RGSD
predictions above 30 LCs, $\sigma_\mathrm{th.}$ must be increased to 1~MeV,
suggesting that the RGSD matrix elements are inconsistent with experimental
constraints. This motivates retaining all 66 LCs for improved SM interactions, a
hypothesis explored in the next section.

This section demonstrated that the EC emulator is an efficient and accurate
substitute for the full SM in fitting interaction matrix elements. Using the
emulator, we successfully reproduced USDA and USDB matrix elements, accelerating
the fitting procedure by a factor of $\approx 800$. These findings pave the way
for constructing novel interactions in regions where SM calculations are
computationally prohibitive, and thus where global fits have not been attempted.

\subsection{LC Markov chain Monte Carlo}\label{sec:mcmc}

In this section, we introduce three new interactions: USDBUQ, USD66, and
USDBUQ-500. All three are constructed using the same experimental data corpus as
USDA/B and the linear combinations identified in Section~\ref{sec:LC}, with $N_d
= 56$ for USDBUQ and $N_d = 66$ for USD66. USDBUQ reproduces the USDB
interaction but incorporates parametric uncertainty quantification (UQ) via MCMC
sampling, providing a more rigorous assessment of uncertainties than the
covariance matrix obtained from \lcmin, which assumes a normal distribution and
neglects higher-order moments.

USD66 generalizes USDBUQ by allowing all 66 linear combinations to vary freely,
treating all matrix elements as independent parameters. While previous
work~\cite{brown2006new} demonstrated that increasing the number of linear
combinations does not significantly improve model accuracy, USD66 is designed to
explore whether the increased uncertainty in the interaction matrix elements
propagates to observables beyond the spectra, such as E2 transitions in
$^{24}$Mg. For USDBUQ, we used a fixed theoretical uncertainty
$\sigma_\mathrm{th.} = 130$~keV, while for USD66, $\sigma_\mathrm{th.}$ was
treated as a hyperparameter and varied alongside the matrix elements, following
the approach in~\cite{pruitt2023uncertainty}.

We then show that both USDBUQ and USD66 underestimate parametric uncertainty,
resulting in predictive intervals that fail to capture the true model error.
This issue, while related to overfitting~\cite{purcell2024improving}, comes from
not accounting for the correlations introduced by the SM itself, and require a
different solution. Instead of reducing the number of degrees of freedom, we
propose increasing the scale of $\sigma_\mathrm{th.}$ to expand the credible
parameter space. This approach leads to USDBUQ-500, which uses
$\sigma_\mathrm{th.} = 500$~keV to recover more realistic confidence intervals.

Table~\ref{tab:interactions} summarizes the new interactions
introduced in this section and compares their performance to the RGSD and USDB
interactions. The interaction samples for USDBUQ, USD66, and USDBUQ-500 are
provided the Supplemental Material~\cite{supp} to facilitate further analysis and
reproducibility.

\begin{table}[htb!]
    \centering
    \begin{tabular}{l c c | c c c c}
         &&& \multicolumn{4}{c}{Standard deviation (keV)} \\
         Interaction & $N_d$ & $\sigma_\text{th.}$ & $x_i-\bar{x}_i$ & $\lambda_{n}-\bar{\lambda}_n$ & $\lambda_{n}-E_n$  & $\bar{\lambda}_n-E_n$  \\
         \hline
         \hline
         RGSD       & 0  & N/A         & - &  - & 1500*   & 1500* \\
         USDB       & 56 & 100       & - &  - & 131*   & 131* \\
         USDBUQ     & 56 & 130       &67 &  45& 141 & 133 \\
         USD66      & 66 & 134$\pm$4 &114&  45& 136 & 133 \\
         USDBUQ-500 & 56 & 500       &201& 134& 189 & 134 \\
         \hline
         \hline
    \end{tabular}
        \caption{Summary of interactions presented in this work as well as
        RGSD~\cite{hjorth-jensen1995realistic} and USDB~\cite{brown2006new}.
        $N_d$ is the number of LCs adjusted
        to data. $\sigma_\text{th.}$ is the theory uncertainty (keV) used in the
        fit. Standard deviations are computed across all posterior samples and either all matrix elements or all energies.
        $x_i-\bar{x}_i$: each interaction matrix element about its mean.
        $\lambda_{n}-\bar{\lambda}_n$: each model energy about its means.
        $\lambda_{n}-E_n$: errors relative to experiment of model predictions. 
        $\bar{\lambda}_n - E_n$: errors relative to experiment of averaged predictions.
        *For RGSD and USDB there is only one prediction per energy $\lambda_n$.
        }
    \label{tab:interactions}
\end{table}

\subsubsection{USDBUQ}

In this section, we present a refit of the USDB interaction using MCMC,
resulting in a new interaction termed USDB with uncertainty quantification
(USDBUQ). The fitting procedure remains largely unchanged, except that the
theoretical uncertainty was increased from 100~keV to 130~keV to better reflect
the empirical DRMS achieved by the original USDB interaction. To enable direct
comparison between the MCMC method and the traditional $\chi^2$ minimization
approach, we used the same set of linear combinations (LCs) obtained in
Section~\ref{sec:LC}. Specifically, we varied the leading $N_d = 56$ linear
combinations of the 66 total matrix elements, while the remaining 10 were fixed
at their RGSD values.

The prior distribution was defined as a multivariate normal with means equal to
the optimal values from the \lcmin fit and standard deviations set to 10 times
the uncertainties derived from the diagonal of the covariance matrix. This broad
prior ensures the algorithm explores alternative solutions while remaining
within the domain of validity of the EC model.

Using the \texttt{emcee}~\cite{foremanmackey2013emcee} affine-invariance
ensemble sampler, we generated 10$^4$ iterations with 560 walkers, yielding a
total of 5.6 million samples. Across all 77 nuclei considered, this corresponds
to over 430 million SM calculations performed using the EC emulator. Posterior
statistics were computed using the final sample from each Markov Chain.

\begin{figure*}[htb!]
    \centering
    \includegraphics[width=\textwidth]{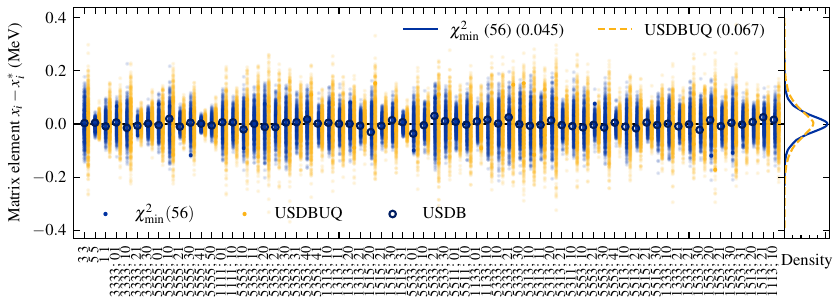}
    \caption{Posterior distribution of new USDBUQ matrix elements (yellow, right
    clusters) relative to the optimal values $x_i^*$ found by $\chi^2$
    minimization in Section~\ref{sec:LC}. The horizontal axis labels each of the
    66 matrix elements with $3=d_{3/2}$, $5=d_{5/2}$, $1=s_{1/2}$ labeling the
    three single-particle orbits. For TBMEs, the labels have the correspondence
    $V_{ijkl; JT} \to ijkl; JT$. Blue (left) clusters show random samples from
    the $\chi^2$ LC method with standard deviations given by the error matrix
    (transformed to matrix elements). Dark blue open-circles show the values for
    USDB. Right subpanel shows the overall normal distributions of matrix
    elements with standard deviations indicated in the legend in MeV.}
    \label{fig:medist-usdbuq}
\end{figure*}

Figure~\ref{fig:medist-usdbuq} shows the marginal distributions of the
single-particle energies (SPEs) and two-body matrix elements (TBMEs) relative to
the optimal matrix elements $\bm{x}^*$ obtained from the \lcmin fit. The
sub-panel highlights the overall variance of the matrix elements for both
methods, as well as the USDB interaction. For comparison, we drew 560
samples from the multivariate normal posterior of the LC $\chi^2_\text{min.}$
fit, with means equal to the optimal LCs $\bm{y}^*$ and variances derived from
the diagonal error matrix $D$. In the LC basis, this multivariate normal
posterior contains no off-diagonal terms.

Each LC sample was transformed into interaction matrix elements using the
transformation $\bm{x} = A^{-1} \bm{y}$. Similarly, the yellow (placed on the
right) clusters in Figure~\ref{fig:medist-usdbuq} represent samples from the
USDBUQ interaction, transformed using the same matrix $A$.

The MCMC results indicate that the mean values of the matrix elements are
consistent with those obtained via $\chi^2_\text{min.}$ fitting. However, the
standard deviation of the matrix elements is larger for USDBUQ, increasing from
approximately 45~keV for \lcmin to 67~keV for USDBUQ. This increase in
uncertainty is directly attributable to the larger theoretical uncertainty
$\sigma_\text{th.} = 130$~keV used in the fit compared to the 100~keV used for
\lcmin. Figure~\ref{fig:medist-usdbuq} further reveals that some matrix
elements, such as the $d_{5/2}$ and $s_{1/2}$ single-particle energies, are
strongly constrained by the data, while others, like the $d_{3/2}$ SPE, remain
less constrained.

In summary, USDBUQ retains the same mean matrix elements as those determined via
LC $\chi^2$-minimization but exhibits uncertainties approximately 50\% larger
due to the increased theoretical uncertainty. The 560 independent walkers
explored the parameter space within ten standard deviations of the \lcmin
minimum, and no alternative minimum was identified. Additionally, the posterior
distribution of linear combinations remained uncorrelated, confirming the
robustness of the linear transformation identified through
$\chi^2$-minimization.

\subsubsection{USD66}

Motivated by the results shown in Fig.~\ref{fig:ylc}, we created another new
interaction, USD66, defined by fitting all 66 linear combinations of the
interaction matrix elements to the data. While this approach provides marginally
better agreement with the data, our primary goal is not to improve the fit but
to avoid arbitrarily fixing the least-constrained linear combinations of the
final 66 matrix elements to their RGSD values. By accounting for all 66 degrees
of freedom, this method allows us to evaluate whether the increased uncertainty
in the matrix elements propagates to other observables not included in the fit.

As with USDBUQ, we use the results of \lcmin to define the set of linear
combinations as an independent set of orthogonal model parameters and to
construct a prior for the 66 parameters. Since our focus now extends beyond
identifying the optimal matrix elements to quantifying their uncertainties and
covariances, we revisit the assumption about the theoretical uncertainty
$\sigma_\mathrm{th.}$. As demonstrated in the case of USDBUQ
(Fig.~\ref{fig:medist-usdbuq}, right panel), increasing $\sigma_\mathrm{th.}$
directly leads to larger uncertainties in the fit results.

We take an additional step by treating $\sigma_\mathrm{th.}$ as a
hyperparameter, allowing the MCMC algorithm to optimize its value. This
approach, as implemented in Ref.~\cite{pruitt2023uncertainty}, incorporates
$\sigma_\mathrm{th.}$ into the log-likelihood function (Eq.~\eqref{eq:logp}),
which includes a term proportional to the determinant of the covariance matrix
($\ln|\Sigma| = \ln\prod_{i=1}^{N_\mathrm{data}}\sigma^2_\mathrm{th.}$). This
term naturally prevents $\sigma_\mathrm{th.}$ from growing indefinitely during
optimization.

\begin{figure*}[htb!]
    \centering
    \includegraphics[width=\textwidth]{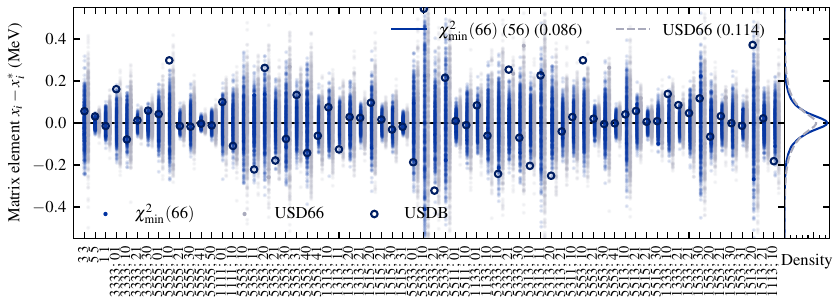}
    \caption{Similar to Fig.~\ref{fig:medist-usdbuq}, distribution of new USD66
    interaction (grey, right clusters) relative to the optimal values $x_i^*$
    found by $\chi^2$ minimization in Section~\ref{sec:LC}. Blue (left) clusters
    show a random sampling from the $\chi^2$ minimization linear combinations
    with standard deviations given by the error matrix. Dark blue open-circles
    show the values for USDB. Right subpanel shows the overall normal
    distributions of matrix elements with standard deviations indicated in the
    legend in MeV.}
    \label{fig:medist-usd66}
\end{figure*}

We generated approximately 10$^4$ MCMC iterations with 670 walkers, yielding a
total of 6.7 million samples. Posterior distribution statistics were computed
using the final sample of the Markov Chain (Fig.~\ref{fig:medist-usd66}). As with
USDBUQ, the uncertainties obtained via MCMC are slightly larger than those
derived from \lcmin. The posterior distribution for $\sigma_\mathrm{th.}$
converged to a value of approximately $134 \pm 4$~keV, which is close to the
130~keV value used in the USDBUQ fit. For comparison, the \lcmin method assumed
$\sigma_\mathrm{th.} = 100$~keV.

The most notable result of the USD66 interaction is the substantially larger
variance observed in some matrix elements compared to USDBUQ. The mean
uncertainty of the matrix elements increased from approximately 86~keV for
\lcmin to 114~keV for MCMC (i.e., the half-widths of the distributions shown in
the subpanel of Fig.~\ref{fig:medist-usd66}). Overall, the mean uncertainty for
USD66 matrix elements (114~keV) is approximately 70\% larger than the
uncertainty reported for USDBUQ (67~keV). This increase is expected, as the
USD66 fit allows all 66 degrees of freedom to vary, whereas 10 of the linear
combinations were held fixed in USDBUQ.

The consequences of this increased matrix element uncertainty for other
observables will be explored in Section~\ref{sec:otherobs}. However, we argue
that reporting the full range of parameter variations compatible with the data
is essential for a complete uncertainty analysis. By accounting for all degrees
of freedom, USD66 provides a more comprehensive representation of the
uncertainties inherent in the interaction parameters.

\subsubsection{Uncertainty estimation and USDBUQ-500}\label{sec:underest}

\begin{figure*}[htb!]
    \centering
    \includegraphics[width=\textwidth]{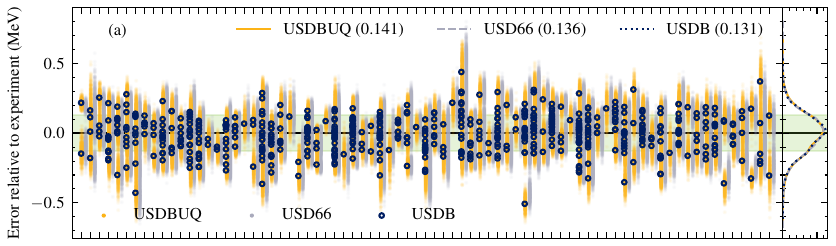}
    \includegraphics[width=\textwidth]{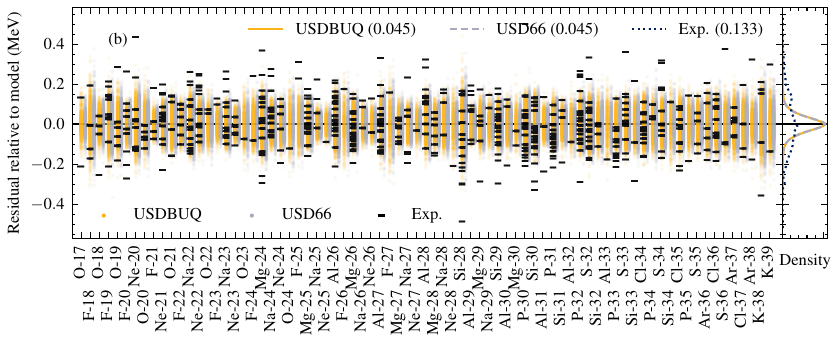}
    \caption{Distribution of residuals relative to a) experimental values and b)
    mean model predictions. Panel a) shows we reproduce the known USDB
    systematic error of 130~keV, while b) demonstrates that the uncertainty
    predicted is a significant underestimation since $\approx40\%$ of
    experimental values fall outside of the 99\% prediction interval. Horizontal
    axis groups the 608 energy predictions by nucleus. Yellow points (left set
    for each nucleus) show USDBUQ predictions and gray points (right set for
    each nucleus) show USD66. Blue, open circles in a) show USDB energies. Black
    dashes in b) show experimental values. The shaded green band in a) indicates
    the $\sigma_\text{th.}=130$~keV value used to fit USDBUQ. Right subpanels
    show the overall normal distributions of a) errors and b) residuals, which
    have means of zero and standard deviations indicated in the legend in MeV.}
    \label{fig:validation}
\end{figure*}

In this section we turn our focus to the more fundamental question of what the
actual uncertainty of a SM prediction is. Ideally, the uncertainty estimated by
sampling uncertainty quantified interactions should reflect the expected
empirical error of the model predictions. However, as illustrated in
Fig.~\ref{fig:validation} (bottom panel), this is not the case using the
assumptions made so far. We demonstrate that the assumption of an uncorrelated
130~keV uncertainty per energy level leads to an underestimation of the
prediction uncertainties. To address this, we introduce our third interaction,
USDBUQ-500, which corrects this underestimation.

Figure~\ref{fig:validation} highlights the performance of the USDBUQ and USD66
interactions in predicting 608 energy levels across 77 nuclei from the training
set. In the main panel, the horizontal axis groups energy levels by nucleus,
while the vertical axis represents the residuals or errors between SM
predictions and either (a) the experimental values or (b) the mean model
predictions. For comparison, the residuals for the original USDB interaction are
shown as dark blue open circles, with one marker per energy level. For example,
$^{39}$K, which has two energy levels in the training set, is represented by two
open circles.

The main results include 560 random samples from the USDBUQ interaction, shown
as yellow clusters on the left for each nucleus, and 670 samples from the USD66
interaction, displayed as gray clusters. The shaded green region in
Fig.~\ref{fig:validation}a) represents the assumed theoretical uncertainty of
$\sigma_\mathrm{th.}=130$~keV used to fit the USDBUQ matrix elements, which is
close to the fitted value of $\sigma_\mathrm{th.}=134\pm4$~keV for USD66.

The small subpanels in Fig.~\ref{fig:validation}~a) and
Fig.~\ref{fig:validation}~b) summarize the overall distribution of errors and
residuals. We found that the distributions of errors for binding
energies and excitation energies were the same, so we plot the combined
distribution. Despite the range of predictions for each energy level in the
USDBUQ interaction, the overall error distribution is comparable to that of the
original USDB interaction. The DRMS for USDB is 131~keV, while for USDBUQ
(across all 560 samples) it is slightly higher at 141~keV (see $\lambda_n-E_n$ in Table~\ref{tab:interactions}.) This increase is due
to the spread in predictions; if the mean energy predictions of USDBUQ are
used (the marginal rather than joint distributions), the DRMS reduces to 133~keV (see $\bar{\lambda}_n-E_n$ in Table~\ref{tab:interactions}.).

The USD66 interaction exhibits a comparable DRMS. For certain nuclei, such as
$^{39}$K, $^{38}$Ar, and $^{17}$O, the increased degrees of freedom in USD66
reduce the residuals, shifting their distributions closer to zero. However, for
other energy levels, the errors increase. Overall, the DRMS decreases from
141~keV for USDBUQ to 136~keV for USD66 across all 608 energy levels.

Figure~\ref{fig:validation}a confirms that the distribution of residuals is
consistent with the assumed theoretical uncertainty of approximately 130~keV.
However, the uncertainty for individual energy levels, as estimated by the
posterior distribution of matrix elements, is significantly smaller. On average,
the standard deviation of model predictions for individual energy levels is only
about 45~keV for both USDBUQ and USD66.

Figure~\ref{fig:validation}b presents the same calculations as
Fig.~\ref{fig:validation}a, but with the origin shifted to the mean model
prediction for each energy level instead of the experimental value. The inset
panel further clarifies the issue: the uncertainty estimated from fitting
interaction matrix elements underestimates the typical empirical error of the
SM. This underestimation persists even when increasing the number of parameters
from 56 (USDBUQ) to 66 (USD66). While the average standard deviation of the
matrix elements increases from 67~keV to 85~keV due to the additional degrees of
freedom, the standard deviation of the predicted energy levels remains unchanged
at approximately 45~keV. Meanwhile, the DRMS, which reflects the empirical
systematic error, decreases only slightly from 141~keV to 136~keV and remains
roughly three times larger than the parametric uncertainty.
These statistics are summarized in the columns $\lambda_n$ and $\lambda_n-E_n$ of Table~\ref{tab:interactions}, respectively.

Parameter estimation using noisy measurements often assumes that the parametric
model is ``perfect.'' Under this assumption, the parametric uncertainty is
interpreted as the total prediction uncertainty. Consequently, as the number of
data points ($N$) increases, the uncertainty in the quantity of interest is
expected to decrease at a rate proportional to $\sim 1/\sqrt{N}$. In theory,
with sufficiently large $N$, the prediction uncertainty would approach zero, a
desirable situation if the model were free of defects or approximations.

However, in practice, the SM includes known approximations and omissions,
meaning that no single parameter set can consistently describe all experimental
data. Furthermore, experimental data often has minimal to negligible
uncertainty. As a certain point, adding more data points to the SM fit does not
improve predictive accuracy~\cite{purcell2024improving}; instead, it leads to
tighter error bars for predictions that increasingly deviate from experimental
observations. This phenomenon is contradictory to what one would ideally expect
from a UQ approach under these conditions, namely, that the prediction error
would remain relatively constant after a certain number of data points is
included in the fit.

To formalize this discussion in terms of the likelihood function, one might
multiply the uncertainties by the reciprocal factor, $\sqrt{N}$, or
(equivalently) multiply the covariance matrix $\Sigma\approx\Sigma_\mathrm{th.}$
by $N$. We found that this approach greatly overestimates the uncertainty, which
is unsurprising since the assumption of $N$ independent degrees of freedom is
simply the opposite extreme (see also Ref.~\cite{pruitt2024role}). 

\begin{figure}
    \centering
    \includegraphics[width=\linewidth]{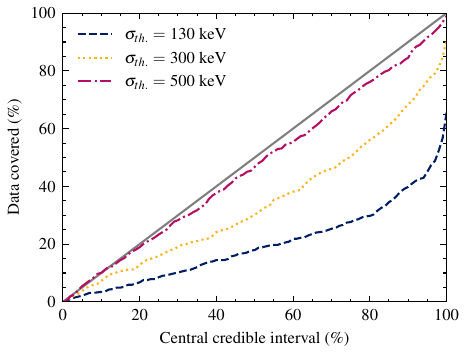}
    \caption{Convergence of the percent experimental data covered as we increase
    the assumed uncorrelated uncertainty $\sigma_\text{th.}$. The uncertainty of
    the predicted energies is smaller than implied by an uncorrelated
    $\sigma_\text{th.}$ due to the off-diagonal terms induced by model
    correlations.}
    \label{fig:coverage}
\end{figure}

To address this situation, we empirically tested the model's coverage and
adjusted the assumed uncorrelated theoretical uncertainty, $\sigma_\text{th.}$,
until a satisfactory agreement between the distribution of model predictions and
the distribution of model errors was achieved (Fig.~\ref{fig:coverage}). As
$\sigma_\mathrm{th.}$ increases, the percentage of experimental data covered by
the central $x$-percent credible interval predicted by the model fit approaches
the ideal 45$^\circ$ line. 

For the USDBUQ and USD66 parameterizations, we initially used
$\sigma_\text{th.}=130$~keV, which corresponds to statistical agreement with the
data under the assumption of $N=608$ independent normal errors. However, to
achieve reasonable coverage of the data, we found it necessary to increase the
assumed uncorrelated uncertainty to $\sigma_\text{th.}=500$~keV. Notably, this
adjustment does not imply that the expected theoretical uncertainty across all
predictions is uniformly 500~keV. Instead, it reflects the need to account for
correlated residuals and model limitations when estimating uncertainties.

\begin{figure*}[htb!]
    \centering
    \includegraphics[width=\textwidth]{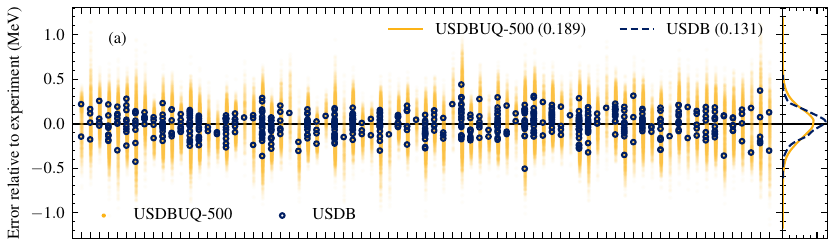}
    \includegraphics[width=\textwidth]{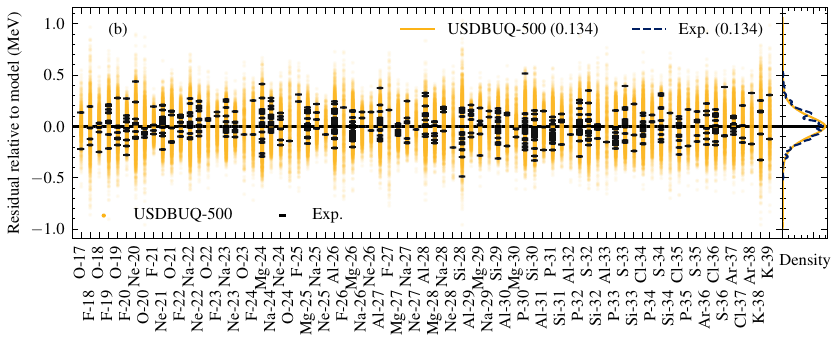}
    \caption{Same as Fig.~\ref{fig:validation} but for the USDBUQ-500
    interaction. Shows the model predictions relative to a) experimental values
    and b) mean model predictions. Using the ``fictional'' uncorrelated
    theoretical uncertainty of 500~keV, the USDBUQ-500 predicted uncertainty
    coincides with the actual standard error. Yellow clusters show the
    distribution of predictions from USDBUQ-500 relative to the mean prediction
    from either a) experiment or b) theory. Black dashes in b) show the
    experimental values relative to the mean model predictions. Right subpanels
    show the overall normal distributions of a) errors and b) residuals, which
    have means of zero and standard deviations indicated in the legend in MeV.}
    \label{fig:validation3}
\end{figure*}

Figure~\ref{fig:validation3} shows the performance of the final interaction
presented in this paper, USDBUQ-500, which was fit identically to USDBUQ but
with $\sigma_\text{th.}=500$~keV. In this case, the standard deviations for both
the model predictions and the residuals of experimental values relative to the
mean model prediction are both 134~keV. The overall distribution of residuals in
both cases is roughly normal with mean zero. However, due to the same
correlations that cause $\sigma_\text{th.}=130$~keV to underestimate the model
uncertainty, $\sigma_\text{th.}=500$~keV does not result in a mean error as
large as 500~keV. Instead, using $\sigma_\text{th.}=500$~keV, the DRMS of the
entire distribution of prediction across the $N=608$ energies increases to only
189~keV (up from 135~keV using $\sigma_\text{th.}=130$~keV). The error of the
mean model prediction using this increased theoretical uncertainty remains
approximately unchanged, increasing from 133~keV to 134~keV. 

To summarize the properties of USDBUQ-500: a single random interaction drawn
from the USDBUQ-500 interaction has an expected DRMS of 189~keV; a large number
of samples drawn from USDBUQ-500 will produce a spread in predictions with an
average half-width of about 134~keV per level (averaged across energy levels for
many nuclei), and the experimental value likewise will be on average 134~keV
away from the mean prediction. These statistics are summarized in Table~\ref{tab:interactions}.
We suggest that USDBUQ-500 provides a more
credible uncertainty analysis than previous attempts.

Finally, to confirm the importance of model covariances, we examined the
covariance matrix of the posterior distribution of energies. As expected, none
of the posterior covariance matrices are found to be diagonal. For USDBUQ, the
posterior covariance matrix $\Sigma_\text{post}$ has 56 singular values (or,
principal components), equal to the number of model parameters. The average of
the USDBUQ singular values is (128~keV)$^2$, which reflects the diagonal
$\Sigma^2_\text{th.}=130^2$~keV$^2$ values used in the fit. On the other hand,
the average of the diagonal elements of the posterior covariance matrix for
USDBUQ ($\Sigma_\text{post}(\text{USBUQ}))$ is only (37~keV)$^2$, which explains
the small variance in energies displayed in Fig.~\ref{fig:validation}. The large
discrepancy between the diagonals and the singular values of
$\Sigma_\text{post}$ shows the large impact of off-diagonal covariance.
$\Sigma_\text{post}(\text{USDBUQ-500})$ also has 56 singular values. The average
of the USDBUQ-500 singular values is (430~keV)$^2$. To the point, the average
diagonal of $\Sigma_\text{post}(\text{USBUQ-500})$ is (128~keV)$^2$, which
correctly captures the empirical uncertainty we set out to reproduce.

\subsubsection{Monopole energies and quadrupole transitions}\label{sec:otherobs}

Next, we compare specific features of the interactions, the single-particle
energies and monopole energies. Furthermore, we wish to test whether the
increased parametric uncertainty described in the USD66 interaction results in
greater uncertainty in transition quantities. 

The single-particle energies $\epsilon_a$, with $a=\{3=d_{3/2},5=d_{5/2},
1=s_{1/2}\}$, are a key part of the interaction as they represent the dominant
mean-field interaction among all the nucleons. The monopole energies are sums of
those two-body matrix elements which can be recast as a number operator, and
therefore contribute to the effective single-particle energies. They are defined
as:
\begin{equation}\label{eq:monopoles}
    M_{ab; T} = \frac{\sum_J (2J+1) V_{ab,ab; JT}}{\sum_J (2J+1)}.
\end{equation}
The effective single-particle energies (ESPEs) are then given by
$\tilde{\epsilon}_a \equiv \epsilon_a + \sum_T M_{aa; T}$, and we report the
mean and standard deviations from each fit (along with USDB) in
Table~\ref{tab:spes} . 

\begin{table}[htb]
    \centering
    \begin{tabular}{l r r r r}
        SPE/ESPE & USDB & USDBUQ & USD66 & USDBUQ500 \\
        \hline
        \hline
        $\epsilon(d_{3/2})$       &  2.112 & $2.11(8)$ & $2.08(13)$ & $2.12(26)$\\
        $\epsilon(d_{5/2})$       & -3.926 & $-3.93(2)$ & $-3.95(2)$ & $-3.92(7)$\\
        $\epsilon(s_{1/2})$       & -3.208 & $-3.20(5)$ & $-3.19(5)$ & $-3.18(15)$\\
        $\tilde\epsilon(d_{3/2})$ & -3.289 & $-3.30(11)$ & $-3.32(15)$ & $-3.29(37)$\\
        $\tilde\epsilon(d_{5/2})$ & -6.897 & $-6.89(4)$ & $-6.98(4)$ & $-6.87(13)$\\
        $\tilde\epsilon(s_{1/2})$ & -6.851 & $-6.84(5)$ & $-6.86(5)$ & $-6.82(16)$\\ 
        \hline
        \hline
    \end{tabular}
    \caption{Single particle energies (SPE, $\epsilon$) and effective single
    particle energies (ESPE, $\tilde\epsilon$) of the new interactions in MeV.}
    \label{tab:spes}
\end{table}

An important lesson from comparing USDBUQ and USD66 single-particle energies is
that increasing the degrees of freedom in USD66 does not significantly increase the
uncertainty (width) of the predicted single-particle energies. However, it does
shift the minimum of the $\tilde\epsilon(d_{5/2})$ single-particle energy.
When the uncertainty is appropriately increased to account for shell model (SM)
defects (as in the USDBUQ-500 interaction), this difference between the minima
becomes insignificant at the 1$\sigma$ level. For USDBUQ-500, uncertainties are
roughly 2–3 times larger, consistent with the increased uncertainty in energy
levels.

We also compare predictions for electric quadrupole $B(E2)$ transition
strengths between low-lying states in $^{24}$Mg, using standard effective
charges $e_n = 0.45$ and $e_p = 1.36$~\cite{richter2008sdshell}. The transitions
analyzed include three from the ground-state rotational band ($6^+_1 \to 4^+_1$,
$4^+_1 \to 2^+_1$, $2^+_1 \to 0^+_1$) and one from a second rotational band
($4^+_2 \to 2^+_2$). Figure~\ref{fig:mg24e2} shows the $B(E2)$ values for these
transitions, and Table~\ref{tab:mg24e2} summarizes the theoretical predictions
compared to experiment.

\begin{figure*}[htb]
    \centering
    \includegraphics[height=.55\textheight]{eval-24Mg-500-Be2}
    \caption{Posterior distribution of the $B(E2)$ values for $^{24}$Mg using
    the new interactions. The right-most set of panels in each row
    show the marginal posterior distribution of each of the four transition
    strengths. The remaining panels show the joint distribution of pairs of
    transitions, with contour lines showing 68\% joint credibility region. As
    with energy levels, USDBUQ-500 predicts a much larger uncertainty in $B(E2)$
    values than USDBUQ or USD66, however it maintains the same correlations
    between transitions within a rotational band.} \label{fig:mg24e2}
\end{figure*}

Both USDBUQ and USD66 interactions predict nearly identical means and
uncertainties for the transition strengths, except for the ground-state
transition, which shows a slight splitting between the two interactions. The
increased degrees of freedom in USD66 do not significantly affect the
uncertainty in the predicted transition matrix elements. In contrast, the
USDBUQ-500 interaction, with its three-fold enhanced interaction uncertainty,
predicts similar mean transition strengths but with uncertainties increased by a
factor of 4–6. This adjustment brings the theoretical uncertainties to
approximately $1–3\%$, though the predictions still fail to agree with
experimental values. 

All four transitions are systematically underestimated by $\approx 50\%$,
suggesting that the standard effective charges underestimate the increased
collectivity that may come from particle-hole excitations that lie outside the
$sd$-shell space. This conclusion is further supported by the strong positive
correlations observed within transitions of the same rotational band
(Fig.~\ref{fig:mg24e2}). Correcting one transition to match its experimental
value would also bring the others closer to their observed values due to these
correlations.

\begin{table}[htb]
    \centering
    \begin{tabular}{c c c c c}
        $i\to f$ & Exp (w.u.) & USDBUQ & USD66 & USDBUQ-500 \\
        \hline
        \hline
        $6^+_1\to 4^+_1$ & $38.0^{+1.8}_{-1.0}$ & $22.9\pm 0.12$ & $22.9\pm 0.12$ &  $22.7 \pm 0.77$\\
        $4^+_1\to 2^+_1$ & $35.7^{+3.4}_{-2.9}$ & $24.9\pm 0.08$ & $24.8\pm 0.10$ &  $24.7 \pm 0.55$\\
        $2^+_1\to 0^+_1$ & $21.07^{+.48}_{-.46}$ & $19.0\pm0.04$  & $18.9 \pm 0.04$  & $19.0 \pm 0.22$\\
        $4^+_2\to 2^+_2$ & $14.9 \pm 1.2$ & $9.8\pm 0.03$ & $9.8 \pm 0.03$ &  $9.8 \pm 0.13$ \\
        \hline
        \hline
    \end{tabular}
    \caption{Reduced matrix elements B(E2) for transitions in $^{24}$Mg shown in
    Fig.~\ref{fig:mg24e2}.} 
    \label{tab:mg24e2}
\end{table}

\section{Conclusions}
\label{sec:conclusion}

In this paper, we demonstrated that eigenvector continuation (EC) can
effectively emulate the SM with sufficient accuracy to fit SM interactions with
negligible error. The speedup achieved by EC scales with the cube of the
dimension for the largest fitted system, resulting in a roughly
three-order-of-magnitude improvement for the \textit{sd}-shell. This
acceleration makes it possible to use MCMC as the fitting algorithm, which we
employed to confirm that the existing USDB interaction with 56 degrees of
freedom represents a global, albeit shallow, minimum. Importantly, once the EC
emulator is constructed, its runtime is independent of the model space size,
enabling its application to more computationally demanding studies in the
future.

Our analysis showed that the interaction distribution obtained via MCMC closely
matches the distribution derived from the \lcmin method. This indicates that
there are no significant nonlinearities between the matrix elements and the
energy spectra, which could otherwise lead to non-normal distributions in the
interaction matrix elements. This finding suggests that the \lcmin approach is
robust and appropriate for future studies.

Regarding uncertainty analysis, we found that restricting the fitting procedure
to fewer linear combinations (e.g., $N_d = 56$ for USDB) underestimates the
total parametric uncertainty of the matrix elements. However, this limitation
has minimal impact on the uncertainty of the predicted energy levels or $E2$
transition probabilities. More importantly, we found that the assumption of
uncorrelated uncertainties, inherent to previous methods, leads to a factor of
three underestimation of the matrix element uncertainties. Consequently, any
confidence intervals derived from such underestimated parameter uncertainties
will systematically underrepresent the true model error.

Approximating the covariance matrix as diagonal introduces an unaccounted
reduction in the uncertainty propagated to the model parameters. To address this
issue, we empirically adjusted the diagonal elements of the assumed covariance
matrix until the predicted model uncertainty matched the observed model error.
Using this approach, we developed the uncertainty-quantified USDBUQ-500
interaction. Sampling from this interaction’s matrix element distribution yields
predicted energy levels with an expected mean error of 0~keV and a standard
deviation of 190~keV (Fig.~\ref{fig:validation3}a). The confidence intervals
derived from these samples accurately represent the probability of experimental
values falling within those intervals (Fig.~\ref{fig:coverage}). Furthermore,
the average prediction for an energy level is expected to have a standard error
of 134~keV, and the expected standard deviation of predictions for that same
level are expected to match at 134~keV (Fig.~\ref{fig:validation3}b).

\begin{acknowledgments}
Prepared by LLNL under Contract DE-AC52-07NA27344 with support from LDRD Project
No.~24-ERD-023. We are grateful for insightful conversations with Cole Pruitt
and Simone Perrotta regarding uncertainty quantification and model defects.
\end{acknowledgments}

\bibliography{biblio}

\end{document}